\documentstyle[prl,preprint,aps,epsfig]{revtex}
\begin{document}                
\draft
\title{Suppression of soft nuclear bremsstrahlung in proton-nucleus
collisions}
\author{
M.J.~van Goethem$^1$,
L.~Aphecetche$^{2,a}$,
J.C.S.~Bacelar$^1$,
H.~Delagrange$^{2,a}$,
J.~D\'{\i}az$^3$,
D.~d'Enterria$^{2,a}$,
M.~Hoefman$^1$,
R.~Holzmann$^4$,
H.~Huisman$^1$,
N.~Kalantar--Nayestanaki$^1$,
A.~Kugler$^5$,
H.~L\"ohner$^1$,
G.~Mart\'{\i}nez$^{2,a}$,
J.G.~Messchendorp$^1$,
R.W.~Ostendorf$^1$,
S. Schadmand$^{1,b}$,
R.H.~Siemssen$^{1}$,
R.S.~Simon$^4$,
Y.~Schutz$^{2,a}$,
R.~Turrisi$^{1,c}$,
M.~Volkerts$^1$,
V.~Wagner$^5$,
and H.W.~Wilschut$^1$
}

\address{$^1$ Kernfysisch Versneller Instituut, Zernikelaan 25,
NL-9747 AA Groningen, The~Netherlands}
\address{$^2$ Grand Acc\'el\'erateur National  d'Ions Lourds, F-14076 Caen
Cedex 5, France}
\address{$^3$ Institut de F\'{\i}sica Corpuscular, E-46100 Burjassot, Spain}
\address{$^4$ Gesellschaft f\"ur Schwerionenforschung, D-64291 Darmstadt,
Germany}
\address{$^5$ Nuclear Physics Institute, 25068 \v{R}e\v{z} u Prahy,
Czech Republic}
\date{\today}
\maketitle
\begin{abstract}
Photon energy spectra up to the kinematic limit
have been measured in 190 MeV proton reactions with light and heavy nuclei
to investigate the influence of the multiple-scattering
process on the photon production.
Relative to the predictions of models based on a quasi-free production
mechanism a strong suppression of bremsstrahlung is observed in the low-energy
region of the photon spectrum.
We attribute this effect
to the interference of photon amplitudes due to multiple scattering of
nucleons in the nuclear medium.
\end{abstract}
In collisions between nucleons electromagnetic radiation can be emitted due
to the rapid change in the nucleon velocity (bremsstrahlung). Accordingly, in
nucleus-nucleus collisions bremsstrahlung is emitted due to the individual
collisions of the constituent nucleons.
Earlier experiments with protons and heavy ions \cite{Nif89}
indicated that bremsstrahlung is dominantly produced in first-chance
proton-neutron collisions. Consequently, dynamical
nuclear reaction models include photon production in the incoherent
quasi-free collision limit, i.e., free nucleon-nucleon (NN) bremsstrahlung cross
sections are employed assuming on-shell nucleons and the intensities of the
individual scattering processes are added rather than their amplitudes.

At sufficiently high energies multiple-scattering processes become
important in reactions of protons with nuclei.
A significant effect on the radiation process is expected due to the
influence of multiple scattering and off-shell propagation of nucleons.
In this case, the bremsstrahlung amplitudes from different steps in the
scattering process interfere and, therefore, the
individual bremsstrahlung contributions may not be added incoherently.
This so called LPM effect was predicted by Landau, Pomeran\v{c}uk, and Migdal
\cite{LPM}
for the successive Coulomb scattering of electrons in matter, resulting in a
reduced bremsstrahlung rate once the mean free path is shorter than the
coherence length.
This suppression has been reported for pair creation from cosmic-ray photons
\cite{Benisz}, and for
bremsstrahlung from high-energy electrons \cite{Klein} in accelerator
experiments.
The general importance of coherence effects on particle production
and absorption in (non-)equilibrium dense matter has been discussed in the
literature
\cite{Knoll_Vosk_PLAP,Kosh99,Lex}. Such effects are, e.g., relevant
for soft photon and dilepton production in hot hadronic matter \cite{Shuryak,Cleymans}.
However, no quantitative analysis for the LPM effect in nuclear
bremsstrahlung has been reported so far.

To study the influence of the nuclear medium on the bremsstrahlung spectrum
we have measured the energy spectra and angular
distributions of photons up to the kinematic limit in reactions of 190 MeV
protons with a range of targets. A strong suppression of bremsstrahlung
relative to a quasi-free production model is observed in the low-energy
regime of the photon spectrum.
The present study was part of the experimental program with the photon
spectrometer TAPS \cite{Str96,Gab94} at the AGOR facility of the
KVI Groningen. A proton beam with a typical intensity of 0.5 -- 2 nA
was incident on solid targets of Au, Ag, Ni, and C with thicknesses ranging
from 20 to 56 mg/cm$^2$. External conversion of photons was kept below 1\% by
the use of a 70 cm diameter carbon-fibre scattering chamber with 4 mm wall
thickness.

The photon spectrometer TAPS was configured in 6 blocks of 64 BaF$_2$ crystals
each at a distance of 66 cm from the target. The setup covered the polar
angular range between 57$^\circ$ and 176$^\circ$ on both sides of the beam with
an azimuthal acceptance of $-21^{\circ} < \phi < 21^{\circ}$. The granularity
of the TAPS setup resulted in an angular resolution of 5.2$^\circ$.
Photons were separated from nuclear particles via their  time-of-flight with
respect to the radiofrequency signal (RF) of the cyclotron.
The time resolution was about 1 ns (FWHM).
In addition, pulse-shape discrimination was employed.
The event trigger required an energy deposition of at least 5 MeV in
a BaF$_2$ module. The signals from the plastic veto detectors in front of the
BaF$_2$ scintillators were used to select photons and protons on the trigger
level.
The relative energy calibration was determined from the characteristic energy
deposited by cosmic-ray muons. The absolute calibration was provided by the
$\pi^0$ mass peak and the 15.1 MeV photons originating
from inelastic proton scattering on $^{12}$C. A small residual
background from cosmic-ray muons within the trigger gate was removed by
subtracting the photon spectrum
obtained by gating on a random time window with respect to the RF.

Two-photon invariant mass spectra from events with two coincident photons
were analysed in order to obtain the $\pi^0$ decay contribution.
The raw $\pi^0$ distributions were
corrected for the finite acceptance and the response of TAPS. The measured
pion distribution was extrapolated by Monte-Carlo simulations
\cite{vanGoethem,Laurent} into regions of missing acceptance by
analyzing the angular distributions in small energy bins of 2 MeV.
The spectrum of photons from $\pi^0$ decay peaks at about 70 MeV with a
yield at least a factor 5 below the inclusive photon yield.
After correction for contributions from $\pi^0$ decay, photon spectra
at laboratory polar angles of 75$^{\circ}$, 115$^{\circ}$ and 155$^{\circ}$ in
a window of $\pm5^{\circ}$ are obtained with a systematic uncertainty of 5\%.
Uncertainties due to beam current and target thickness contribute another 5\%.
Here we present the results near 90$^{\circ}$ in order to minimize the
influence of the reference frame.

\begin{figure}
\centerline{\epsfig{file=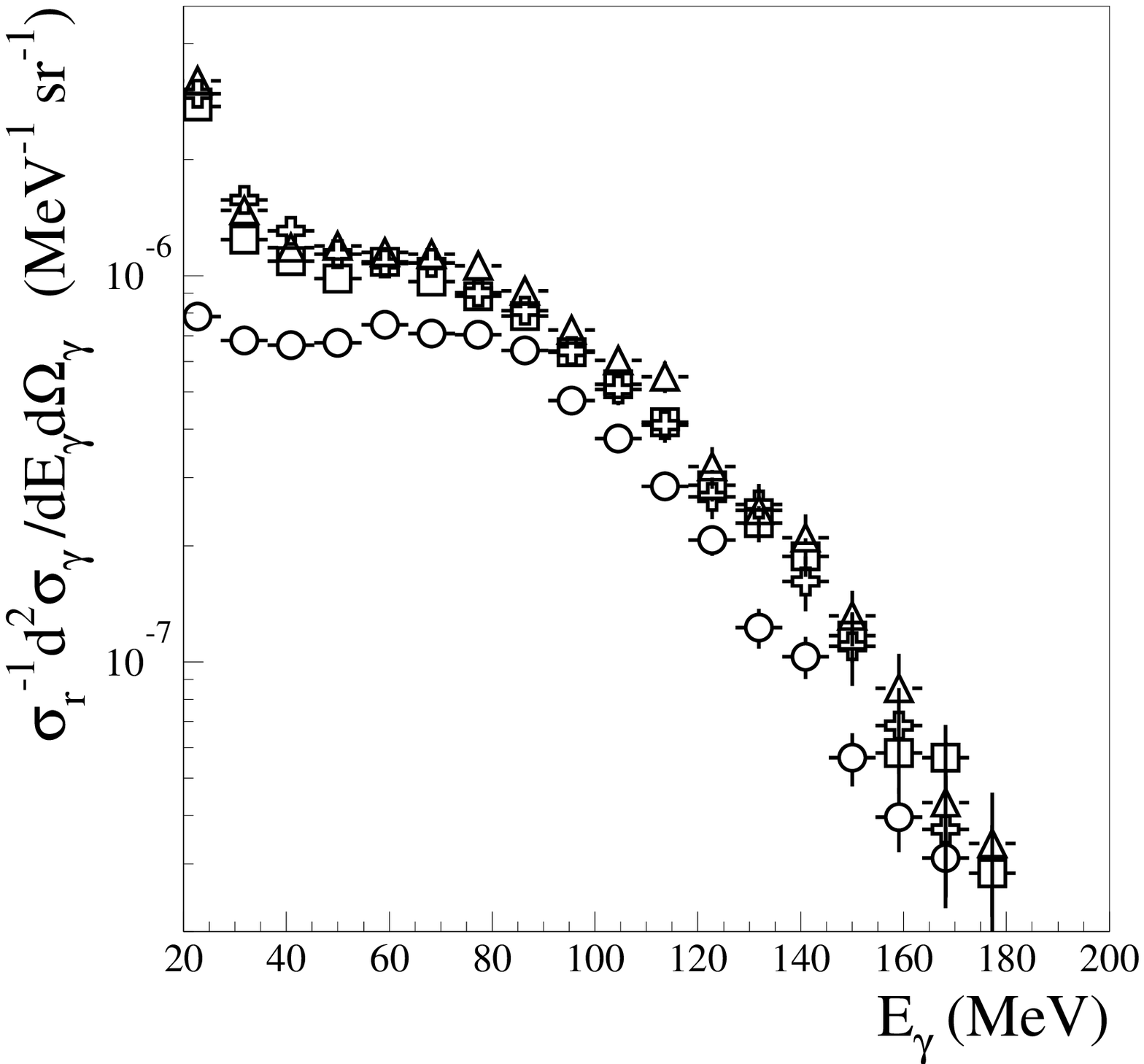,width=7.cm}}
\caption{Target-mass dependence of photon spectra for 190 MeV protons on
C (circles), Ni (triangles), Ag (squares), and Au (crosses) targets at a lab.
angle of 75$^{\circ}$. The double differential cross sections have been
normalized to the geometrical reaction cross section.}
\label{massdependence}
\end{figure}

Fig.~\ref{massdependence} shows a compilation of the photon spectra
at a lab. angle of 75$^{\circ}$ for the four targets studied here.
The double differential cross sections have been normalized to the
geometrical cross section $\sigma_r=1.44\pi A^{2/3}$ fm$^2$ of
each reaction with target mass number $A$.
The spectra extend up to the kinematic limit $E_{max}$=$T_{CM}$+Q, where Q is
the Q-value of the reaction and $T_{CM}$ the center-of-mass energy.
If plotted as function of the scaled photon energy
$E_{\gamma}$/$E_{max}$, all data in Fig.~\ref{massdependence} above 100 MeV
fall on the same curve \cite{vanGoethem,Kwato}.
The shape of the photon spectra displays a plateau
between 30 and 80 MeV and an exponential decrease
towards the kinematic limit. This shape is different from photon spectra in
heavy ion reactions \cite{Schutz,Martinez}, where nearly exponential slopes have
been observed above 30 MeV. The rise at photon energies below 30 MeV for
the heavier targets can be attributed to statistical photon emission.

For comparison with dynamical model
calculations including the multiple-scattering process we employ the
Intra-Nuclear Cascade (INC) code of Cugnon \cite{Cugnon}. The INC
model was chosen because it reproduces well many aspects of proton-nucleus
reactions at these bombarding energies \cite{Cugnon-pA}.
Furthermore it allows the study of photon-nucleon correlations \cite{MJcorr}
because it conserves correlations between scattered particles.
This is a necessary feature for the inclusion of the NN bremsstrahlung
production in a non-perturbative manner with the kinematically correct
$pn\gamma$ process using the Soft-Photon Approximation (SPA) for free
$pn$ bremsstrahlung \cite{Alex_Olaf}.
We found that the optimal way to implement Pauli blocking in INC was
achieved by requiring all final scattering states to lie above the Fermi
surface. The nucleon phase-space distributions from the INC calculations 
agree well with those obtained using the Boltzmann-Uehling-Uhlenbeck 
(BUU) \cite{Cassing} transport model. The BUU results describe the photon 
spectrum ($E_\gamma > 30$ MeV) from 180$A$ MeV Ar+Ca collisions fairly 
well \cite{Martinez} although slightly overpredicting the yield on the 
soft side of the spectrum.

\begin{figure}
\centerline{\epsfig{file=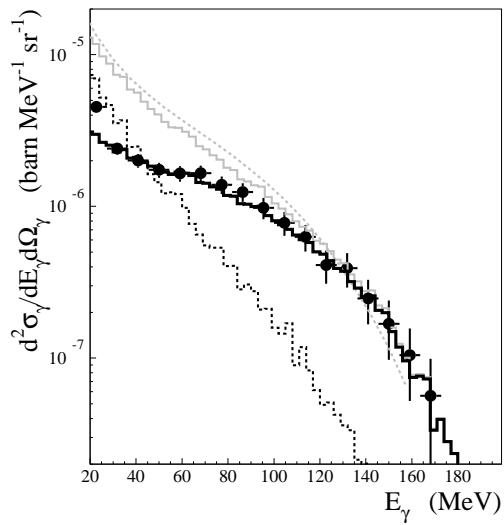,width=7.cm}}
\caption{Photon spectrum for 190 MeV p+Au at an angle of
$75^{\circ}$ (filled circles), compared to results of the INC
(grey histogram) and BUU (dashed line) models.
The dashed histogram shows the multiple-step contribution from the INC model.
The black histogram is the INC result multiplied with the
quenching factor $f_q$ from Eq.\ (\protect{\ref{genqf}}).}
\label{inclusive}
\end{figure}

Fig.~\ref{inclusive}
shows the photon spectrum for the Au target at
75$^{\circ}$ in comparison with the INC and BUU results.
Both models agree quite well with each other in the full spectrum but
overestimate significantly the experimental photon yield at low and
intermediate photon energies (E$_{\gamma}\leq$100 MeV).
Good agreement between the experimental data and the calculations
is found in the hard part of the spectrum, even near the kinematic limit.
The multiple-step contribution from INC has been indicated separately in
Fig.~\ref{inclusive}. The overall agreement between experiment and theory
would be much better if no multiple scattering was taken into account.
It seems as if multiple-scattering processes are overestimated in theory.
However, the amount of multiple scattering in INC
has been checked against the experimental proton yields at large angles.
These proton yields, in which multiple-scattering
processes are essential, agree within the error margins for all targets
studied \cite{vanGoethem}. Therefore, multiple scattering is well described.

In the nuclear medium the nucleon mean free path is
$\lambda_{mfp} =1/(\rho\cdot\sigma_{NN}) \approx 2$ fm,
based on an average NN scattering cross section $\sigma_{NN} = 30$ mb
\cite{Franz} at 190 MeV and the nuclear saturation density
$\rho = 0.16$ fm$^{-3}$. Therefore, nuclear bremsstrahlung can be quenched for
a photon wavelength $\lambda \ge \lambda_{mfp}$ or a photon energy
$E_\gamma \le \hbar c/\lambda_{mfp} \approx 90 $ MeV. The strength of quenching
of course increases with decreasing photon energy.
 In a simplified model based on the classical description of bremsstrahlung
production in hard collisions we have estimated the analytical shape
 of the LPM effect in a two-step $p$+nucleus reaction \cite{vanGoethem}.
 Each segment of the proton trajectories defines a
 production amplitude with a definite relative phase and therefore must be
 added coherently.
The time between two collisions is characterized by the mean collision time
$\tau = \lambda_{mfp}/(g \beta_0 c) = \tau_0/g$ with
$\beta_0$ the incoming proton velocity, i.e.\ $\tau_0\approx 4$ fm/c. The 
factor $g$ takes into account that
in subsequent collisions the velocity of the leading particle is reduced. A
value $g\approx 0.5$ is expected to describe the mean time between the first
and second collision, i.e., $\tau \approx 8$ fm/c.
Averaging over the time
distribution $(1/\tau)exp(-t/\tau)$ we derive the following quenching factor,
whose analytical form is motivated by several theoretical calculations
\cite{Knoll_Vosk_PLAP,Cleymans,Alm}:
\begin{equation}
f_q=\xi\left(1-\frac{\alpha}
  {1+(\frac{E_\gamma}{\hbar}\tau)^2}\right).
  \label{genqf}
\end{equation}
The parameter $\alpha$ is related to the fraction of energy remaining for the
leading particle in subsequent collisions \cite{vanGoethem}, i.e.,
$\alpha \approx 1/g^2 \approx 0.25$. $\xi$ is an overall scaling factor.
The INC calculation was adjusted in an ad hoc manner to account for medium
effects by multiplying the spectrum obtained from INC with the quenching factor
$f_q$ from Eq. (\ref{genqf}).

The experimental data were fitted with the product $INC \cdot$ $f_q$,
where $INC$ represents the full INC spectrum and $\alpha$, $\xi$ and $\tau$
are free parameters. For the spectra obtained
at three different angles of 75$^{\circ}$, 115$^{\circ}$ and 155$^{\circ}$
we find
$\alpha\approx 1$ and a mean collision time
$\tau=2.4\pm 0.6$ fm/c for the Ni, Ag and Au targets
and $3.7\pm 0.5$ fm/c for the C target.
These values for the collision time are much smaller than the expected
average time interval
between two hard NN collisions in nuclei (8 fm/c, see above).
From this observation one must conclude that hard NN collisions
alone are insufficient to explain the quenching of soft photons.
Other effects likely to increase the observed collision frequency (reduced
parameter $\tau$) may be
multiple soft collisions, but also
a modification of the elementary photon production process in the nuclear
medium. The latter hypothesis
is supported by the observation that the dipole contribution expected from
the elementary proton-neutron angular distribution appears to be absent in
the reactions studied here, as was observed also elsewhere \cite{Clayton}.
In our data we observe the corresponding result
from the fact that  $\xi$ increases from $\xi=0.96\pm 0.06$
at $75^\circ$ to $\xi= 4.3\pm 2.0$ at $155^\circ$.

\begin{figure}
\centerline{\epsfig{file=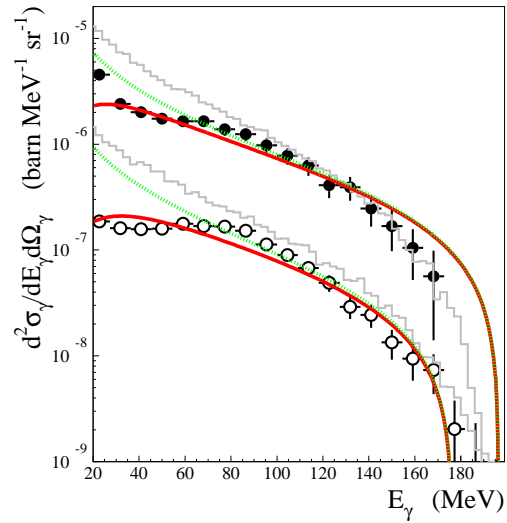,width=7.cm}}
\caption{Photon spectrum at a lab. angle of
$75^{\circ}$ for 190 MeV p+Au (top, filled circles) and
p+C (bottom, circles). The grey histograms indicate the
results from the INC model.
The dark lines are obtained using Eq.\ (\protect{\ref{genfE}}), the dotted
grey lines correspond to Eq.\ (\protect{\ref{genfE}}) with the collision 
frequency set to zero.}
\label{fig4ab}
\end{figure}

The separation between dynamical effects and LPM quenching is complicated
due to the partitioning of the NN interaction in the nuclear medium into a
mean-field and a collision component in the models.
The available dynamical models are all of semiclassical nature and our
new data indicate the need to include consistently the medium
modifications and the interference phenomena.
The development of a quantum transport theory for photon production in
intermediate-energy proton+nucleus
reactions, was already started in ref.\ \cite{Nakayama} but did not go
beyond the conventional
quasi-particle approximation, i.e., the correlations and off-shell nucleon
propagation in the medium were not
taken into account. A new approach has been taken up recently \cite{Lex}.
The in-medium photon production cross section is calculated from a microscopic
NN interaction including the spectral width of the
baryon propagators.
Two sources of multiple-scattering effects can be
identified: one of minor importance is scattering of final state nucleons
before or after photon radiation; more significant appears to be the multiple
scattering during off-shell nucleon propagation before the photon is emitted.
Preliminary results \cite{Armen} for a 200 MeV $p$+nucleus reaction
indicate that this approach leads to a remarkable suppression of soft photons
below 80 MeV, in qualitative agreement with our data. This approach
requires, however, a non-zero temperature of the target nucleons.
Recently, another theoretical approach \cite{Koshelkin} was taken
where the kinetic equations that determine the evolution of the two-particle 
Green's function in matter were derived
in the transport approximation for soft-photon production. The correlations 
in the medium allow multiple scattering to occur without requiring multiple 
hard collisions, thus yielding Eq.\ (\ref{genqf}) with
$f_q(\alpha=1, \tau=\tau_0)$, i.e. $g=1$, in agreement with the empirical 
result. We thus obtain an energy dependence of the photon spectrum with the 
functional form
\begin{equation}
f_E \sim \frac{E_\gamma}{E_\gamma^2 +(\hbar/\tau_0)^2(Z/N)^2}\cdot
\beta_0^2\cdot (1-E_{\gamma}/E_{max})
  \label{genfE}
\end{equation}
This spectrum incorporates the factor $1-E_{\gamma}/E_{max}$ to describe the
kinematic limit which is absent in the soft photon approach.
(The precise form of this limit may vary, cf.\ \cite{Clayton}.)
The proton to neutron ratio (Z/N) describes the reduced quenching observed in
heavy nuclei. The absolute cross section
is determined by the geometrical reaction cross section.
The suppression of soft photons
can be described quantitatively as shown in Fig.~\ref{fig4ab} by the
dark lines. The amount of quenching can be seen from comparison with the
result where the collision frequency (i.e., $\hbar/\tau_0$) is set to zero
(the grey dotted lines).  This approaches the quasi-free result of
the INC model. 
Eq.\ (\ref{genfE}) describes also well the published data
\cite{Kwato,Clayton,olddata}
for 168 MeV p+Tb and 145 MeV p+Pb by only adjusting $\beta_0$ according to the
respective beam energy.
This shows that also at lower beam energy quenching occurs, but the effect at 
photon energies above 30 MeV is small and went unnoticed so far.

In summary, new experimental data have been presented for nuclear
brems\-strah\-lung from
the soft-photon region up to the kinematic limit in proton+nucleus reactions.
We observe a strong suppression of the soft brems\-strah\-lung cross section in
comparison with the prediction of transport models that include bremsstrahlung
on basis of quasi-free nucleon-nucleon collisions.
Applying a phenomenological quenching factor appropriate for sequential
hard  collisions between nucleons, we can fit the data using the average
collision time as a free parameter. We find that its value is much shorter
than expected on basis of the collision times in a transport model.
New theoretical models are being developed using simplified reaction dynamics
but taking into account medium effects such as nucleon correlations and 
off-shell propagation of the nucleons involved in the photon production.
An analytical form for the bremsstrahlung production in nuclear matter based on
these assumptions was  obtained which describes the suppression of soft 
bremsstrahlung. These results show the importance of multiple-scattering 
processes beyond the classical picture of multiple hard collisions.

{\it The effort of the AGOR team in providing high-quality beam is
gratefully acknowledged. We thank the KVI theory group and A. Korchin, 
A.V. Koshelkin, and A. Sedrakian for valuable discussions and for providing 
results from their calculations prior to publication.
This work was supported in part by FOM, the Netherlands, by IN2P3 and
CEA, France, BMBF and DFG, Germany, DGICYT and the Generalitat
Val\`encia, Spain, by GACR, Czech Republic,
and by the European Union HCM network contract HRXCT94066.}

\end{document}